\documentclass[UKenglish,twocolumn,amsmath,amssymb, prb, aps, floatfix,superscriptaddress, 10pt]{revtex4-2}

\DeclareUnicodeCharacter{0308}{}

\usepackage{listings}
\usepackage{color}
\usepackage{graphicx}
\usepackage{caption}
\usepackage{physics}
\usepackage{braket}
\usepackage{ulem}
\usepackage{hyperref}
\hypersetup{colorlinks}

\usepackage{xcolor}
\usepackage{tabularray}

\definecolor{myblue1}{rgb}{0.29, 0.58, 0.92}
\definecolor{myblue2}{rgb}{0.5, 0.71, 0.94}
\definecolor{myblue3}{rgb}{0.64, 0.8, 0.96}
\definecolor{myblue4}{rgb}{0.79, 0.89, 0.98}
\definecolor{myblue5}{rgb}{0.92, 0.96, 0.99}

\definecolor{mygreen1}{rgb}{0.85, 0.96, 0.96}
\definecolor{mygreen2}{rgb}{0.71, 0.95, 0.92}

\definecolor{myyellow1}{rgb}{0.99, 0.88, 0.53}
\definecolor{myyellow2}{rgb}{0.97, 0.76, 0.14}

\newcommand{\cnn}{Universit\'e Paris-Saclay, Centre de Nanosciences et de Nanotechnologies, CNRS, 10 Boulevard Thomas Gobert, 91120, Palaiseau, France}
\newcommand{\qdl}{Quandela SAS, 7 Rue L\'eonard de Vinci, 91300 Massy, France}
\newcommand{\sydney}{School of Mathematical and Physical Sciences, University of Technology Sydney, Ultimo, New South Wales 2007, Australia}
\newcommand{\imperial}{Department of Physics, Imperial College London, London, UK}
\newcommand{\pariscite}{Universit\'e Paris Cit\'e, Centre de Nanosciences et de Nanotechnologies, CNRS, 10 Boulevard Thomas Gobert, 91120, Palaiseau, France}

\begin{document}

\title{Indistinguishability of remote quantum dot-cavity single-photon sources}

\author{Mathias Pont}
\email{mathias.pont@quandela.com}
\affiliation{\cnn}
\affiliation{\qdl}
\author{Stephen C. Wein}
\affiliation{\qdl}
\author{Ilse Maillette de Buy Wenniger}
\affiliation{\cnn}
\affiliation{\imperial}
\author{Valentin Guichard}
\affiliation{\cnn}
\author{Nathan Coste}
\affiliation{\cnn}
\affiliation{\sydney}
\author{Abdelmounaim Harouri}
\affiliation{\cnn}
\author{Aristide Lema\^itre}
\affiliation{\cnn}
\author{Isabelle Sagnes}
\affiliation{\cnn}
\author{Lo\"ic Lanco}
\affiliation{\cnn}
\affiliation{\pariscite}
\author {Nadia Belabas}
\affiliation{\cnn}
\author{Niccolo Somaschi}
\affiliation{\qdl}
\author{Sarah E. Thomas}
\affiliation{\cnn}
\affiliation{\imperial}
\author{Pascale Senellart}
\email{pascale.senellart-mardon@cnrs.fr}
\affiliation{\cnn}

\begin{abstract}
     Generating identical photons from remote emitter-based bright single-photon sources is an important step for scaling up optical quantum technologies. Here, we  study the  Hong-Ou-Mandel interference  of photons emitted from remote  sources based on semiconductor quantum dots.  We make use of a deterministic fabrication technique to position the quantum dots in a spectrally resonant micropillar cavity and fine tune their operation wavelength  electrically. Doing so, we can match four pairs of sources between five distinct sources, study them under various excitation schemes and measure their degree of indistinguishability. We demonstrate remote indistinguishabiltiy between $44\pm1$\%  and $69\pm1$\% depending on the pair of sources and excitation conditions,  record values  for quantum dots in cavities. The relative contribution of pure dephasing and spectral diffusion  is then analysed, revealing that  the remaining distinguishability is mostly due to low frequency noise.
\end{abstract}

\date{\today}
\maketitle

\section{Introduction}

The number of qubits involved in  photon-based quantum information processing protocols is largely determined by the efficiency of indistinguishable single-photon sources. Two main single-photon source technologies are now routinely exploited. Sources based on photon pair generation in non-linear media and photon heralding allowed impressive progresses~\cite{zhong2018, wang2018, vigliar2021}. Due to the probabilistic  generation process, a massive integration of sources and detectors is required to scale up~\cite{PSQ2024}. More recently, single-photon sources (SPS) based on semiconductor quantum dots (QDs) have been shown to generate on-demand indistinguishable single-photons, at high efficiency when inserted in optical cavities~\cite{senellart2017, somaschi2016, tomm2021, uppu2020}. Their compatibility with low-loss integrated optical circuits, and high-efficiency single-photon detectors have made them important contenders for intermediate scale quantum computation \cite{wang2019, cao2023, hansen2023, maring2023}. To generate multi-photon states with QD-based sources, the current approach has been to use active demultiplexing of a temporal stream of identical photons generated by one bright source~\cite{wang2019, li2020, li2021, pont2022, pont2022GHZ, maring2023, cao2023}. In parallel to increasing a single source efficiency, further scaling will eventually require making use of many identical SPS~\cite{wein2024}. 

The generation of indistinguishable photons from remote QD based single-photon sources has motivated many works in the past fifteen years~\cite{patel2010, flagg2010, gao2013, he2013, gold2014, giesz2015, kim2016, jons2017, reindl2017, weber2018, thoma2017, zopf2018, weber2019, zhai2021, you2022, papon2022, dusanowski2023}. The first requirement for high remote source indistinguishability is the ability to tune the two sources to the same wavelength within the emission linewidth. The two photonic wavepackets should also present the same temporal profile, a challenging requirement when the QDs are inserted into an optical cavity for high photon collection:  both emitters should then undergo similar acceleration of spontaneous emission. Last but not least, each quantum emitter should emit  infinite trains of identical photons. This requires that the QD does not suffer from  decoherence during the emission process (no pure dephasing) nor on a long time scale (no spectral diffusion). 
An important milestone has recently been achieved with the demonstration of a remote two-photon indistinguishability of ($93.0\pm0.8$)\% with GaAs QDs in a bulk medium~\cite{zhai2021}, albeit at low collection efficiency. Very few works have studied the indistinguishability of remote QD in cavities - ensuring high collection efficiency, with  remote indistinguishability at best in the $40\ \%$ range~\cite{giesz2015}.

Here, we report on the  measurement of remote two-photon indistinguishability from multiple QD-cavity based single-photon sources. We harness a deterministic fabrication technique based on optical in-situ lithography~\cite{dousse2008} to fabricate two samples with multiple QD-micropillar sources. Selection of QDs with similar as-grown emission wavelength combined with fine electrical spectral tuning allow us to identify five sources distributed in four pairs of spectrally matched sources presenting a classical temporal overlap between $(98.4\pm0.1)\%$ and $(99.6\pm0.1)\%$. We study the quantum interference of photons emitted by these various remote sources under either resonant or phonon-assisted excitation schemes. All pairs  reach record values of remote two-photon indistinguishability for QD in cavities up to $69\pm1\%$. Time delay dependent analysis of the indistinguishability of successively emitted photons from the most identical sources allow us to identify the contribution of pure dephasing and spectral wandering to the remaining distinguishability.
 
\section{Single photon source devices}

\begin{figure}[h]
    \centering  
    \includegraphics[width=0.98\linewidth]{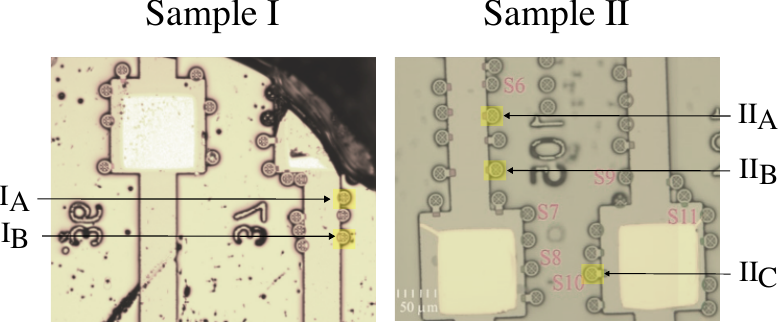}
    \caption{Optical microscope image of the two samples under study. We indicate on each the sources that can be brought to resonance with another source of the other sample. The four pairs   are \{I$_A$, II$_A$\}, \{I$_A$, II$_B$\}, \{I$_B$, II$_B$\} and \{I$_B$, II$_B$\}.}
    \label{fig:samples}
\end{figure}

We study two samples labelled I and II, shown in Fig.~\ref{fig:samples}, fabricated from the same molecular-beam-epitaxy grown planar cavity wafer consisting of isolated InGaAs QDs embedded in a $\lambda$-cavity with 14 (28) GaAs/AlAs Bragg pairs in the top (bottom) mirror. A 20 nm-thick Ga$_{0.1}$Al$_{0.9}$As barrier, positioned 10 nm above the QD layer, is used to increase the hole capture time inside the QD. The \textit{in-situ} lithography technique~\cite{dousse2008, nowak2014, somaschi2016} was used to deterministically position a single QD within 50 nm of the center of a connected pillar cavity.  The Purcell effect in the weak coupling regime allows for a large fraction of the single-photon emission to be funnelled into the pillar cavity mode.  The in-situ lithography process allows the adjustment of the pillar cavity diameter to ensure the spectral resonance between the QD and the cavity lines is better than 0.5~nm~\cite{dousse2008}.

Each sample embeds a vertical p-i-n diode structure and the pillars are electrically contacted to operate in the reverse-bias regime. This enables different occupation states for the QD ground state depending on the applied voltage. In the following, we study both neutral and charged QD sources. The electrical bias also allows us to fine tune the QD-cavity detuning within a 0.3-0.5~nm range depending on the source~\cite{nowak2014}.
The performances of the samples studied here were previously benchmarked in terms of source brightness, single-photon purity and indistinguishability as reported in Ollivier et al.~\cite{ollivier2020}. An average indistinguishability of $(90.6 \pm 2.8)\%$, single-photon purity of $(95.4 \pm 1.5)\%$ were reported over 5 samples, including samples I and II. The average first lens brightness $\mathcal{B}$, i.e. the probability to get a photon per excitation pulse at the output of the micropillar device,  was found to be $(13.6\pm4.4)\%$.

The mean operation wavelength yielding optimal performances of all the sources across samples I and II was shown to be 924.7~nm with a standard deviation 0.5~nm. Among the single-photon sources of each samples, we identify several pairs of sources that can be tuned to the same emission energy. Fig.~\ref{fig:samples} shows optical microscope images of the two samples and highlight the 5 sources, two on sample I (I$_A$ and I$_B)$, three  on sample II (II$_A$, II$_B$, II$_C$) that can be spectrally matched two by two to form 4 pairs as shown later on.   

We explore the interference of photons emitted by remote sources using two excitation schemes: resonant excitation and longitudinal acoustic (LA) phonon  assisted excitation. Both excitation techniques have been shown to lead to high indistinguishability for photons successively emitted by the same source~\cite{Thomas2021}. Resonant excitation results in a near unity occupation probability of the QD excited state that allowed record source brightness~\cite{tomm2021,ding2023high}. However, it  requires to collect the single photons in crossed polarisation with  respect to the excitation laser polarisation~\cite{he2013,somaschi2016,tomm2021} which remove useful spin information for graph state generation. LA-phonon-assisted excitation allows to spectrally filter the laser and to  generate polarization-entangled multi-photon states through spin-photon entanglement~\cite{coste2023high}, at the expense of a slightly reduced excited state occupation probability. Both schemes  lead to different temporal profiles, which influences the remote source interference.

\section{Wavelength and temporal overlap}

High indistinguishability for single photons generated by two quantum emitter sources  first requires that both  sources are spectrally matched and present the same emission profile. For two remote dipoles labelled $i$ and $j$ showing mono-exponential decay times $T_{1,i}$ and  $T_{1,j}$, in the absence of any decoherence process, the two-photon mean wavepacket overlap is given by \cite{giesz2015}

\begin{equation}
\label{eq:M-T1}
    M_{i,j}=\frac{4\gamma_i\gamma_j}{(\gamma_i+\gamma_j)^2+\Delta_{i,j}^2},
\end{equation}

\noindent with the emission rate defined as $\gamma_{i}=1/T_{1,i}$ and $\gamma_{j}=1/T_{1,j}$ and $\Delta_{i,j}=\omega_i-\omega_j$ the frequency difference between the two sources. When the two sources are spectrally matched ($\Delta_{i,j}=0$),  the upper limit for remote indistinguishability is then the classical temporal overlap $s_{i,j}=4\gamma_i\gamma_j/(\gamma_i+\gamma_j)^2$ which reaches unity for $\gamma_i=\gamma_j$. 

\noindent For emitters in cavities, the  lifetime depends on the detuning between the emission energy of the QD and the energy of the cavity mode. Moreover, the highest photon collection efficiency for each source requires  each QD transition to be tuned to their respective cavity resonance. Hence, a strong requirement for optimal temporal and spectral overlap is to fabricate devices with strongly overlapping cavity resonances.

\begin{figure}[t]
    \centering  
    \includegraphics[width=\linewidth]{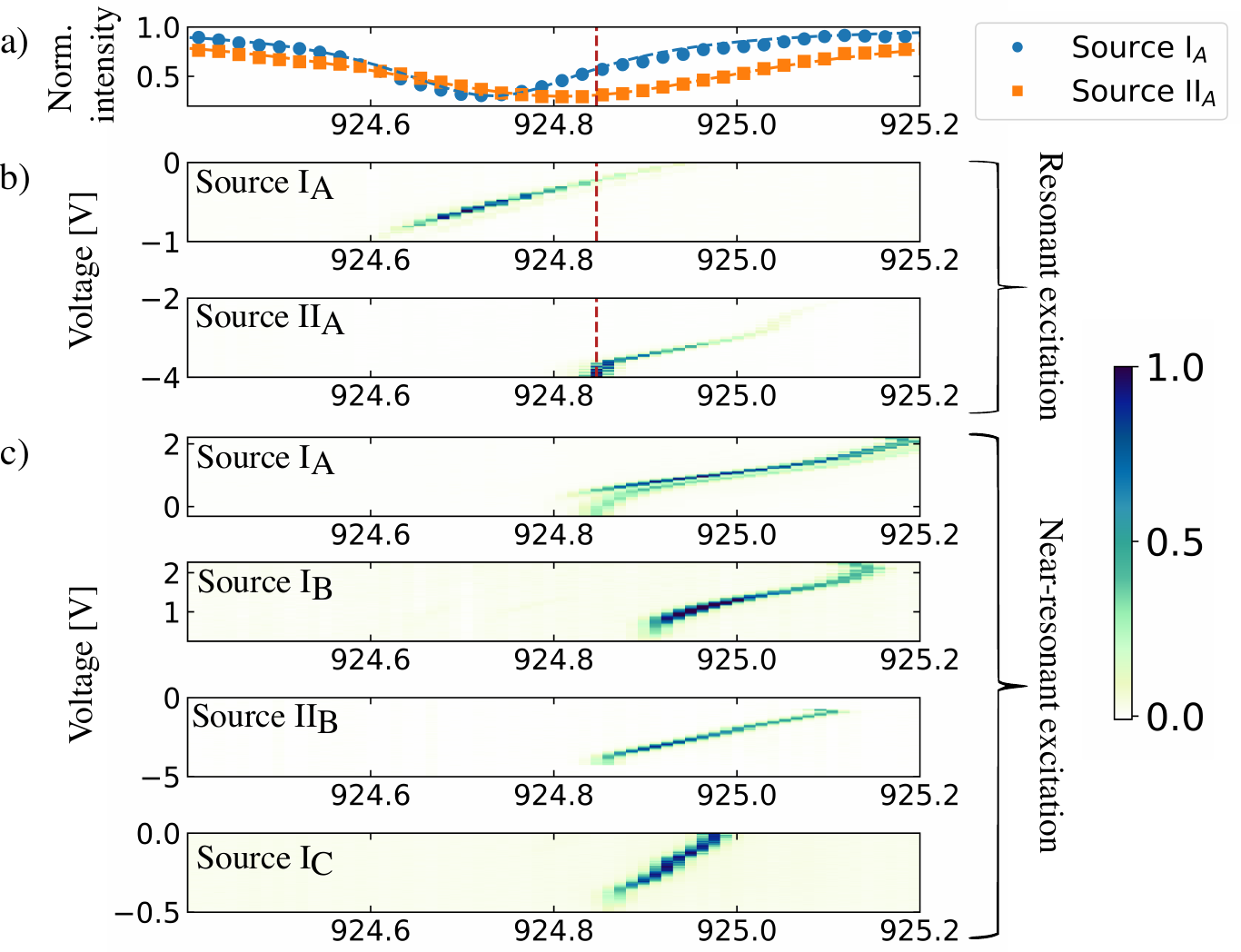}
    \caption{(a) Reflectivity spectra of source I$_A$ and II$_A$ measured using a broad-band LED source in a polarisation resolved reflectivity setup. (b-c) Emission wavelength of each source as a function of the applied voltage bias under (b) resonant excitation and (c) LA-phonon-assisted excitation exictation. As shown in (a-b) with the mathching pair I$_A$ and II$_A$, by tuning two sources into resonance (indicated by the dotted red vertical line) we move away from the exact individual resonances of the microcavities described by Lorentzian fits for both source I$_A$ (blue circles) and source II$_A$ (orange squares) shown in panel (a).}
    \label{fig:tuning}
\end{figure}

Fig.~\ref{fig:tuning}a shows the reflectivity spectra of two devices ($I_A$ and $II_A$), one on each sample, measured using a broadband LED source in a polarisation resolved reflectivity setup. The measurement evidences substantial overlap of the two cavity modes and the lorentzian cavity fit yields Q$_{I_A}$=2900 (Q$_{II_A}$=1700) at x$_{c,I_A}$=924.734~nm (x$_{c,II_A}$=924.817~nm) for source I$_A$ (II$_A$) where the quality factor Q$_i$ is the ratio of the central frequency x$_{c,i}$ and the full width at half-maximum of the fundamental mode of the micro-cavity of source $i$.

Fig.~\ref{fig:tuning}b-c show the QD emission energy of each source tuned by application of a  bias. The current flowing through the diodes are measured to be below pA. We use both for resonant excitation (b) and LA-phonon-assisted excitation (c). 
The intensity maps  allow to define the voltage range where  each source can be put in resonance with another one. We  identify 4 pairs of sources that can be tuned to respective resonance: (\{I$_A$, II$_A$\}, \{I$_A$, II$_B$\}, \{I$_A$, II$_C$\} and \{I$_B$, II$_B$\}). For most pairs, under LA-phonon-assisted excitation, we can set a voltage where both sources are closed to mutual QD energy resonance while being close to their maximal emission intensity.

\noindent For resonant excitation,  the same $\sim15$~ps laser pulse is used to excite both QDs so that the scan does not reflect the full tuning range of the source but is constrained by the chosen laser energy. We note moreover that the use of much lower pump power in resonance excitation (below 1~nW) compared to LA-phonon-assisted excitation (few $\mu$W), displaces the bias tuning range of interest.  Fig.~\ref{fig:tuning}b presents a scan under resonant excitation for sources I$_A$ and II$_A$. We observe that while I$_A$ is tunable over a wide energy range,  QD II$_A$ can not be tuned below 924.847~nm (Fig.~\ref{fig:tuning}b). In this case, we  tune the emission energy of I$_A$  to the wavelength where II$_A$ is the brightest (indicated by the dotted red vertical line).  The detuning between source I$_A$ (source II$_A$) and its microcavity is thus $\Delta_{I_A}$=113~pm ($\Delta_{II_A}$=30~pm). This mutual resonance point leads to a  decrease of the brightness for I$_A$ from $\mathcal{B}_{I_A}\sim15.4~\%$ to $\mathcal{B}_{I_A}\sim12.4~\%$, while the brightness of II$_A$ is $\mathcal{B}_{II_A}\sim18.5~\%$.

\begin{figure}[t]
    \centering
    \includegraphics[width=\linewidth]{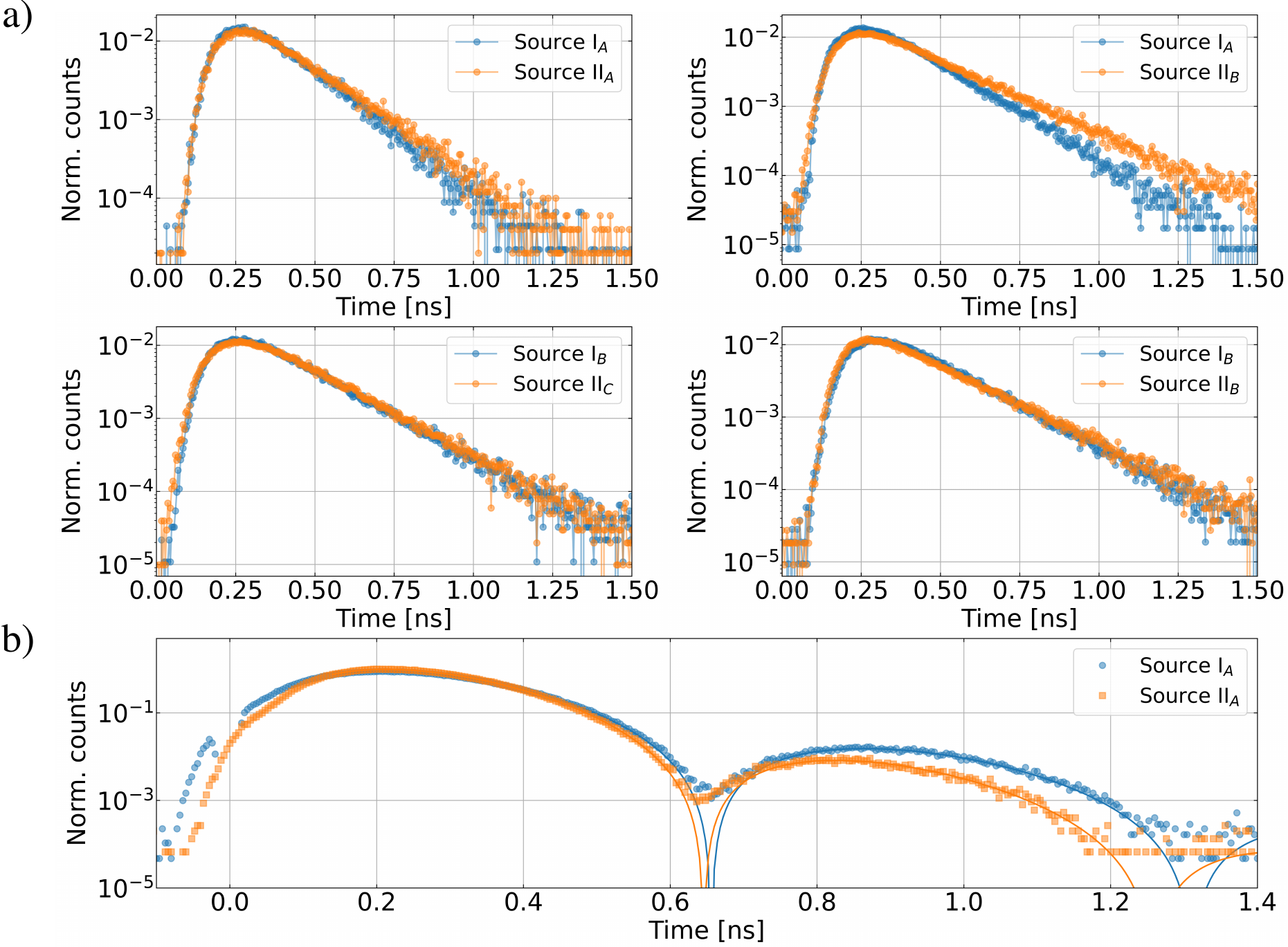}
    \caption{Overlapped time evolution of the PL emission (a) under LA-phonon-assisted excitation of source \{I$_A$, II$_A$\}, \{I$_A$, II$_B$\}, \{I$_B$, II$_B$\} and \{I$_B$, II$_B$\}, and (b) resonant PL emission for source \{I$_A$, II$_A$\}. The integrated counts per time bin have been normalised by the integrated area of the full temporal profile.}
    \label{fig:lifetime}
\end{figure}

We now turn to the temporal profiles of emission of the sources at their mutual resonance. Here, the use of phonon-assisted excitation or resonant excitation alters significantly the temporal emission profiles. The simple situation described by Eq.~\eqref{eq:M-T1} correctly describes the case of phonon-assisted excitation for which a monoexponential decay emission profile can always be obtained independently of the charge state of the QD.

\noindent A charged QD (trion)  corresponds to a four-level system for which both the ground and the excited states are degenerate. Such state structure can  effectively be described as two-level system  with a single time constant for the emission hence mono-exponential decay. Conversely, a neutral QD is a three-level system composed of a single ground state and two non-degenerate linearly polarised excited states. Using LA-phonon-assisted excitation, one can align the linear polarisation of the excitation along  one of the excitonic dipoles reducing the system to an effective two-level system~\cite{Thomas2021}, hence again to a monoexponentially decaying emission. We can thus use this excitation scheme to obtain a similar temporal profiles between neutral and charged QD sources.  We find four different pairs of sources on samples I and II, that can be tuned in resonance, and have a very similar time evolution profiles with typical decay time $T_{1,i}$ within 128-240~ps range as shown in Fig.~\ref{fig:lifetime}a.  We deduce the corresponding classical temporal overlap $s_{i,j}$ which ranges from $(98.4\pm0.1)\%$ to $(99.6\pm0.1)\%$.

Under strictly resonant excitation, achieving such high  temporal overalap for neutral QD sources is more  demanding. This is illustrated now using sources I$_A$ and II$_A$ that embed neutral QDs with  excited state composed of two non-degenerate states~\cite{gammon1996} labelled $\ket{X}$ and $\ket{Y}$ with energy splitting $\Delta_{FSS}$. For a QD in a cavity under resonant excitation, the laser light polarisation is set along the $\ket{V}$ axis of the micropillar cavity in order to suppress any rotation of the laser polarisation by the cavity birefregence~\cite{ollivier2020}. $\ket{V}$ is not parallel to the axis of the QD dipoles $\ket{X}$, one thus creates a superposition of the QD dipoles $\ket{X}$ and $\ket{Y}$, and observe a beating between those two excited states. Fig.\ref{fig:lifetime}b shows the time evolution of each source measured along the $\ket{H}$ axis of their respective microcavities. 

\noindent To describe such situation that cannot be accounted within Eq.~\eqref{eq:M-T1}, we introduce the generalised classical overlap:
\begin{equation}
    s_{i,j}=\left[\int f_i(t)f_j(t)dt\right]^2, 
\end{equation}
where 
$f_i(t)\propto\sqrt{\braket{\hat{a}_i^\dagger(t)\hat{a}_i(t)}}$ is the magnitude of the temporal amplitude of the emitted single-photon pulse normalised to $\int f_i^2(t)dt= 1$. Importantly, this classical overlap bounds the mean wavepacket overlap (see Supporting Information).
We fit the measured temporal decay  with
\begin{equation}
    \label{eq:lifetime}
    |\braket{H|\Psi(t)}|^2=\sin^2 \left( \frac{\Delta_{FSS}}{2 \hbar}t \right) ~\sin^2 \left( 2\theta \right) ~e^{-\frac{t}{T_{1, X}}}
\end{equation}
\noindent {where $\theta$ is the angle between the optical polarization axis $\ket{H}$ and the dipole orientation of the exciton eigenstate $\ket{X}$}. We find the parameters $T_{1, I_A}=(162\pm1$)~ps, $\Delta_{FSS,I_A}=(6.3\pm0.1)~\mu eV$, $T_{1,II_A}=(128\pm1)$~ps and $\Delta_{FSS, II_A}=(6.7\pm0.1)~\mu eV$. To compute the temporal overlap we use the measured temporal decay and obtain a value as high as $s_{I_A,II_A}=(98.6\pm0.1)\%$ despite the slight discrepancy in fine structure splitting. It is worth noting that such high temporal overlap from natural QDs under resonant excitation is more of a lucky coincidence, since not only do they need to have similar excitonic energy splitting but also similar angles between the polarisation of the QD dipoles and the cavity  axes.

\section{Remote source indistinguishability}
\begin{figure}
    \includegraphics[width=0.95\linewidth]{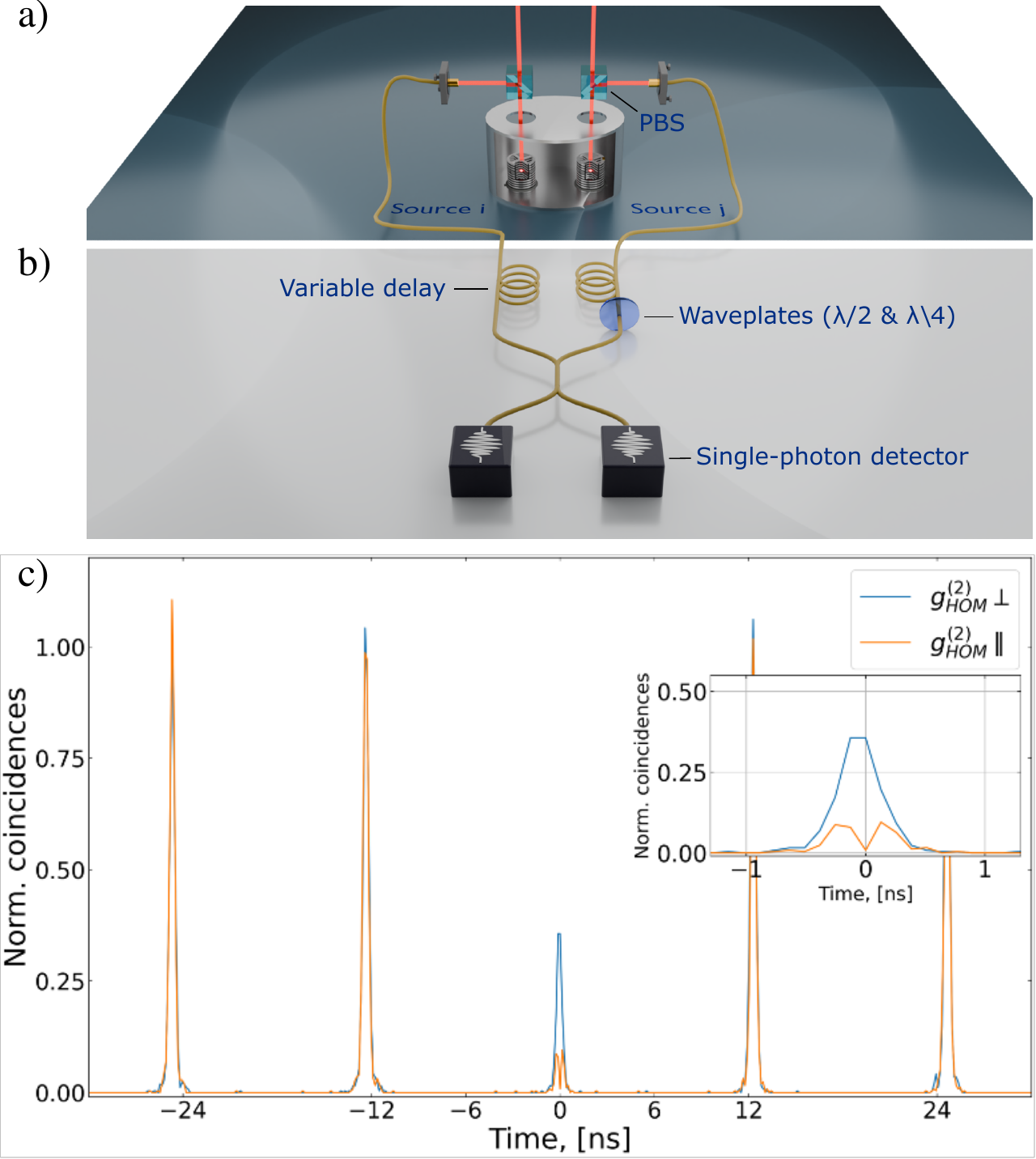}
    \caption{(a) Scheme of the experimental setup used to measure the indistinguishability of photons emitted by two remote SPS (see text) (b) Hong-Ou-Mandel (HOM) interferometer for two-photon interference from remote emitters.  (c) Exemple of correlation measurement acquired for sources $I_A$ and $II_A$ under resonant excitation using parallel ($\parallel$) and perpendicular ($\perp$) polarisation. The measurement gives a HOM visibility of $V_{TPI}=(69\pm1)\%$.} 
    \label{fig:HOM}
\end{figure}

\begin{table*}[ht]
	\centering
    \begin{tblr}{
        colspec = {|c|c|c|c|c|c|c|},
        cell{2,4,6,8}{2} = {myblue1!40},
        cell{3,5}{2} = {myblue2!40},
        cell{7,11}{2} = {myblue3!40},
        cell{9}{2} = {myblue4!40},
        cell{10}{2} = {myblue5!40},
        cell{2-6,8}{3} = {mygreen1!40},
        cell{7,9,10,11}{3} = {mygreen2!40},
        }
		\hline
		Pair Nº & Source label & Charge state & Excitation & $T_{1,i}$ (ps) & $s_{i,j}$ & V$_{TPI}$  \\
		\hline
		\SetCell[r=2]{c}{1} & I$_A$ & X & \SetCell[r=2]{c}{Res.} & 162$\pm$7 & \SetCell[r=2]{c}{0.986$\pm$0.001} & \SetCell[r=2]{c}{0.69$\pm$0.01} \\
		\cline{2-3}\cline{5-5}
		& II$_A$ & X & & 128$\pm$3 & & \\
		\hline
		\SetCell[r=2]{c}{1} & I$_A$ & X & \SetCell[r=2]{c}{LA} & 158$\pm$12 & \SetCell[r=2]{c}{0.992$\pm$0.001} & \SetCell[r=2]{c}{0.463$\pm$0.001} \\
		\cline{2-3}\cline{5-5}
		& II$_A$ & X & & 172$\pm$4 & & \\
		\hline
		\SetCell[r=2]{c}{2} & I$_A$ & X & \SetCell[r=2]{c}{LA} & 145$\pm$10 & \SetCell[r=2]{c}{0.984$\pm$0.001} & \SetCell[r=2]{c}{0.50$\pm$0.01} \\
		\cline{2-3}\cline{5-5}
		& II$_B$ & CX & & 195$\pm$10 & & \\
		\hline
		\SetCell[r=2]{c}{3} & I$_A$ & X & \SetCell[r=2]{c}{LA} & 174$\pm$10 & \SetCell[r=2]{c}{0.996$\pm$0.001} & \SetCell[r=2]{c}{0.44$\pm$0.01} \\
		\cline{2-3}\cline{5-5}
		& II$_C$ &CX & & 219$\pm$10 & & \\
		\hline
		\SetCell[r=2]{c}{4} & I$_B$ & CX & \SetCell[r=2]{c}{LA} & 240$\pm$10 & \SetCell[r=2]{c}{0.995$\pm$0.001} & \SetCell[r=2]{c}{0.59$\pm$0.01} \\
		\cline{2-3}\cline{5-5}
		& II$_B$ & CX & & 212$\pm$10 & & \\
		\hline
	\end{tblr}
	\caption{Remote two-photon indistinguishability $V_{TPI}$ of 4 matching pairs of sources depending on their charge state (neutral QD -- X, charged QD -- CX), the excitation scheme (resonant excitation -- Res., LA-phonon-assisted excitation -- LA), the lifetime $T_{1,i}$ of the excited state, and the generalised classical overlap, $s_{i,j}$. All values are obtained with the additional etalon spectral filter.}
	\label{tab:all_sources}

\end{table*}

We now study the quantum interference of each pair ($i$, $j$) of sources. Fig.~\ref{fig:HOM}a-b present the general scheme for the experimental setup. Both sources are placed in the same closed cycle cryostat operating at $\sim$~5~K on two piezo stages.  A common pulsed excitation laser with a repetition rate of ~82 MHz and pulse duration of $\sim$10~ps is used to excite both sources.   Each source is electrically tuned using two voltage controllers to set  V$_{i}$ and V$_{j}$. The single-photon stream from each source is sent into a fibered Hong-Ou-Mandel interferometer. A fiber delay and a free-space micrometer linear stage are used to ensure that both photons arrive simultaneously at the input of a fibered 50:50 beam-splitter (BS).
Fig.~\ref{fig:HOM}b shows a schematic for the HOM interference where using a set of waveplates we can either set the polarisation of the photons to be parallel ($\parallel$) or orthogonal ($\perp$).  We measure correlations at the output of the beam splitter using high resolution ($\sim$10~ps) superconducting nanowire single-photon detectors and a low time-jitter correlator ($\sim$12~ps).  When both photons are co-polarised ($\parallel$), they interfere at the beam-splitter and we observe a strong antibunching, i.e. a reduction of coincidences at zero time delay. Because the sources under study can show residual blinking, the antibunching peak at zero is normalised with the central peak of the non-interfering histogram when their polarisation in orthogonal ($\perp$) \cite{jons2017, weber2018}. The interference-fringe visibility is thus defined by

\begin{equation}
    V_{TPI} = 1-\frac{A_{\parallel}}{A_{\perp}},
\end{equation}

\noindent where $A_{\parallel}$ and $A_{\perp}$ is the integrated zero-delay peak in presence of quantum interference and without quantum interference, respectively. The interference  of sources I$_A$ and II$_A$ under resonant excitation shows an interference visibility of $V_{TPI}=(54.8\pm1)$\%. Such value already constitutes a record value for remote single photon sources based on QDs in cavities~\cite{giesz2015}. This value is further increased to $V_{TPI}=(69\pm1)$\% (see Fig.~\ref{fig:HOM}c) when the single-photon are filtered with an 8~pm air-spaced etalon at the cost for reducing the effective source brightness by roughly 50\%. This spectral filtering that does not influence the temporal decay of the state (see Supporting Information), as the width of the filter is larger than the radiative linewidth of the QDs. The increased visibility thus indicates some influence of either phonon-sideband emission or spectral wandering i.e. low frequency charge noise. The same pair of sources shows a reduced remote interference visibility of $V_{TPI}=(46.3\pm0.1)\%$ when operated in the phonon-assisted excitation regime.
 
Tab.~\ref{tab:all_sources} presents the experimental data gathered on all the matching pairs on both samples, including excitation scheme, temporal decay time, classical temporal overlap, and remote two-photon indistinguishability. All values reported are measured with the additional (8~pm) spectral filter where very low second-order correlation function $g^{(2)}(0)$ between 0.3\% and 1\% are reached. The remote two-photon indistinguishability ranges from 44\% to 69\% - with an average value of $(54\pm9)$\%, setting new records for QD sources in cavities.

\section{Discussion}

To understand the physical phenomena at the origin of
the remaining distinguishability between remote sources,
we consider the \textit{individual} two-photon indistinguishability $M_i$ of source $i$ for successively emitted photons~\footnote{The single-photon indistinguishability $M$ can be computed
from the visibility of the Hong-Ou-Mandel interference fringe $V_HOM$ and the $g^{(2)}(0)$ as $M=\frac{V_{HOM}+g(^{(2)}(0)}{1-g^{(2)}(0)}$~\cite{ollivier2021}.}. The corresponding values have been measured for the 2 pairs yielding the best \textit{remote} two-photon indistinguishability and are shown in Tab.~\ref{tab:individual_M} for photons emitted 12~ns appart. The measured value above 93.8\% indicate a low degree of pure dephasing for each source.
\begin{table}
	\centering
    \begin{tblr}{
        colspec = {|c|c|c|c|c|},
        cell{2}{1} = {myblue1!40},
        cell{3}{1} = {myblue2!40},
        cell{4}{1} = {myblue4!50},
        cell{5}{1} = {myblue5!50},
        }
		\hline
		Source label & Excitation & M$_{i}$ & $\sqrt{M_iM_j}$\\
		\hline
		I$_A$ & \SetCell[r=2]{c}{Res.} & 93.9$\pm$0.2\% & \SetCell[r=2]{c}{95.5$\pm$0.1\%}\\
        \hline
		II$_A$ & & 97.1$\pm$0.2\% & \\
        \hline
		I$_B$ & \SetCell[r=2]{c}{LA} & 96.6$\pm$0.4\% & \SetCell[r=2]{c}{95.2$\pm$0.1\%} \\
        \hline
		II$_B$ &  & 93.8$\pm$0.2\% & \\
        \hline
	\end{tblr}
	\caption{Individual two-photon indistinguishability of sources \{I$_A$, II$_A$\} for successively emitted photons (12~ns appart) when paired under resonant excitation, and \{I$_B$, II$_B$\} when paired under LA-phonon-assisted excitation.}
	\label{tab:individual_M}
\end{table}

Considering  the influence of pure dephasing only, the expected remote mean wavepacket overlap $M_{i,j}$ is given by

\begin{equation}\label{exponentialMWPO}
M_{i,j} = s_{i,j}\frac{(\Gamma_i+\Gamma_j)(\gamma_i+\gamma_j)}{(\Gamma_i+\Gamma_j)^2+4\Delta_{i,j}^2},
\end{equation}

\noindent where $\Gamma_i=\gamma_i+\gamma^\star_i$ is the total Lorentzian spectral width (see Supporting Information). With the measured near-unity classical temporal overlap $s_{i,j}\sim 1$, and the high degree of indistinguishability between successively emitted photons, we find an upper bound to the remote two photon visibility $V_{TPI}=M_{i,j}$ given by $M_{i,j} \le min \left(s_{i,j}, \sqrt{M_iM_j}\right)$ (see Supporting Information). Note that, we can identify here the remote mean wavepacket overlap $M_{i,j}$ with the two- photon indistinguishability $V_{TPI}$ considering the negligible $g^{(2)}(0)$ of the sources. We find that the measured
values of remote indistinguishability reported in Tab.~\ref{tab:all_sources} do not reach the upper bounds deduced from individual source indistinguishability, showing that it is not limited by temporal, frequency matching or pure dephasing.  

The reduced values of remote indistinguishably is thus mainly limited by $\Delta_{i,j}$ fluctuating over the measurement time scale, i.e. the independent spectral wandering of each source. We make the assumption that each source follows independent normal distributions with mean frequency $\omega_i$ and standard deviations $\delta\omega_i$~\cite{loredo2016}. We define the random variable $\Delta_{i,j}$, also normally distributed, with a mean value $\overline{\Delta}_{i,j}=\omega_i-\omega_j$ and standard deviation $\delta\omega_{i,j}=\sqrt{\delta\omega_i^2+\delta\omega_j^2}$ and average Eq. (\ref{exponentialMWPO})~\footnote{such simplication can be done for lorentzian single photon spectra}.

For the specific case of photons with a Lorentzian spectral shape, we can simply average Eq. (\ref{exponentialMWPO}). Note that this is equivalent to taking the convolution of a Lorentzian function with a Gaussian function, and hence the result of the mean wavepacket overlap contains a Voigt function $V$:
\begin{widetext}
\begin{equation}
\begin{aligned}
    \braket{M_{i,j}}_\Delta &= \frac{\pi}{2}s_{i,j}^2(\gamma_i+\gamma_j) V(\overline{\Delta}_{i,j};\overline{\Gamma}_{i,j},\delta\omega_{i,j})
\end{aligned}
\label{eq:MAB}
\end{equation}
\end{widetext}
where $\overline{\Gamma}_{i,j}=(\Gamma_i+\Gamma_j)/2$ is the average spectral width of the two photons. To test the validity of this analysis, we measure the indistinguishability of different photons emitted by the same source as a function of the time delay between the emitted photons of each source for source I$_A$ and II$_A$ to access the temporal decoherence $\gamma^\star_i$ and the amplitude of the emission energy fluctuation $\delta \omega_i$.

\begin{figure}[h!]
    \centering
    \includegraphics[width=0.95\linewidth]{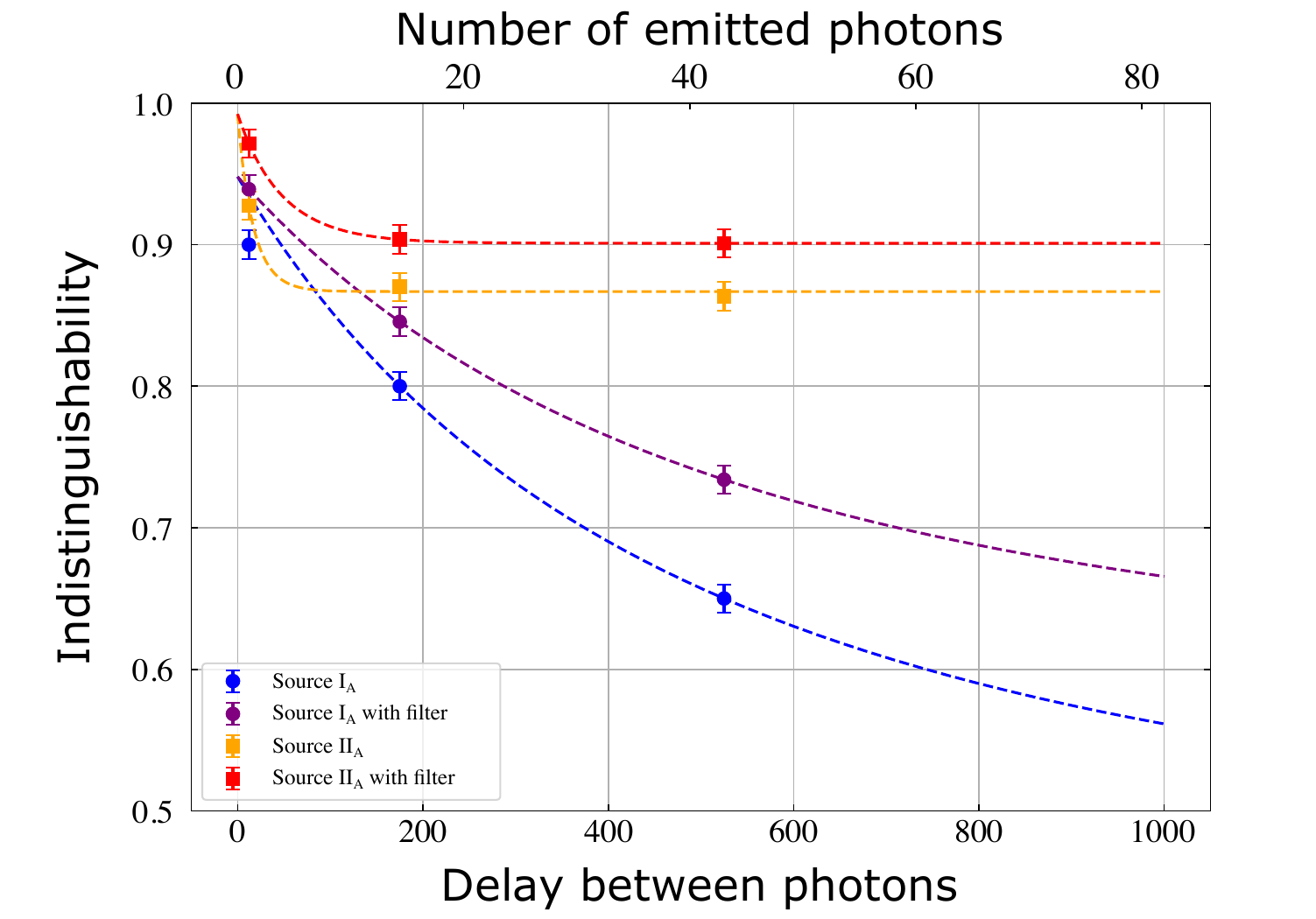}
    \caption{Indistinguishability of sources I$_A$ and II$_A$ as a function of the delay between the 2 interfering photons for a resonant excitation scheme. The spectral filter improves the visibility of the HOM interferences by filtering part of the phonon sideband, and reducing the spectral wandering of the emission line.}
    \label{fig:longldelayhom}
\end{figure} 

A variable delay is set on a path-unbalanced Mach-Zender interferometer to study the indistinguishability of photons separated by an increasing time up to 525 ns. Fig.~\ref{fig:longldelayhom} shows the results obtained for sources I$_A$ and II$_A$ with and without an additional 8 pm width spectral filter. The individual indistinguishability of source I$_A$ decreases from $(89.65\pm0.01)\%$ to $(64.6\pm0.1)\%$ going from a 12.2~ns to a 525~ns delay, while for source II$_A$ the indistinguishability decreases from $(92.75\pm0.01)\%$ to $(86.36\pm0.1)\%$ without spectral filter. With the additional spectral filter, the indistinguishability of source I$_A$ decreases from $(93.9\pm0.2)\%$ to $(73.4\pm0.2)\%$ going from a 12.2~ns to a 525~ns delay, while for source II$_A$ the indistinguishability decreases from $(97.1\pm0.2)\%$ to $(90.1\pm0.2)\%$

Considering a near-perfect single photon purity ($g^{(2)}(0)=0$), the visibility dependence is given by~\cite{loredo2016}:

\begin{equation} \label{eq:visibility_delay}
	V(\tau)=\frac{V(\tau=0)}{1+2\delta\omega_r^2(1-e^{-\Delta_{\tau}/\tau_c})}
\end{equation}

\noindent where $V(\tau=0)=\gamma/(\gamma + \gamma^{\star})$ is the "intrinsic" degree of indistinguishability limited by pure dephasing only, $\tau_c$ a characteristic wandering timescale, and $\delta\omega_{r}=\delta\omega/(\gamma+\gamma^{\star})$ is the ratio between the frequency detuning and the spectral linewidth $\gamma + \gamma^{\star}$.

We fit Eq.~\eqref{eq:visibility_delay} on the dataset presented in Fig.~\ref{fig:longldelayhom} to extract a lower bound the spectral amplitude of spectral diffusion for each source~\footnote{The datapoint for source I$_A$ without filtering for a delay between photons of $\Delta t$=12~ns has been treated differently in the numerical fit as it does not allow the fit to converge. We use an error bar of 10\% during fit to account for a possible outlier, but display the measure value of 2\% in Fig.~\ref{fig:longldelayhom}}. For a single source, we use the same value of indistinguishability $V(0)$ for a zero time delay between photons. For each source,  we can then extract the broadening due to pure dephasing and spectral wandering of both sources using Eq.~\eqref{eq:visibility_delay} to derive $\gamma_{I_A}^{\star}=(0.17\pm0.01)$~ns$^{-1}$ and  $\gamma_{II_A}^{\star}=(0.03\pm0.01)$~ns$^{-1}$ and fit $\delta\omega_{I_A}=4.7$~ns$^{-1}$ ($\delta\omega_{I_A}=4.6$~ns$^{-1}$), and $\delta\omega_{II_A}=2.12$~ns$^{-1}$ ($\delta\omega_{II_A}=1.78$~ns$^{-1}$) without (with) spectral filtering. Using Eq.~\eqref{eq:MAB}, we  derive a limit for $M_{I_A,II_A}$, yielding $M_{I_A,II_A} < (65\pm1)\%$~($M_{I_A,II_A} < (71\pm1)\%$)    without (with) spectral filtering, values that are in good agreement with our remote source interference measurements.

We note that the samples under study show increased spectral diffusion compared to other source samples as reported previously in references~\cite{loredo2016, desantis2017} a difference that could be attributed to a difference in the dopants used in the p-i-n diode. 

\section{Conclusion}

In conclusion, we have reported on record remote indistinguishability for multiple deterministically fabricated quantum dot sources in cavities. Remote indistinguishabilities in the 44-69\% range were reported on several pairs of sources. All measurements were taken  at maximum occupation probability for each source, i.e. at maximal brightness at mutual source resonance. Our study shows that the reduction of  two-photon interference visibility mostly arises from  spectral wandering, most probably originating from slow  electrical  and magnetic noise. More studies will be required to further reduce the influence of these noises. In particular, better understanding could be obtained exploiting well-known noise spectroscopy techniques~\cite{kuhlmann2013} to identify the influence of the as-grown material structure (doping, p-i-n diode structure), the influence of the cavity processing recipes (etching, annealing) as well as the stability of the control electronics. We finally note that these high values of remote indistinguishability have been obtained both on neutral and charged QDs. The remote interference of such bright sources embedding a single spin acting as a local quantum bits has recently been shown to be highly beneficial to scale-up optical quantum computing~\cite{deGliniasty2024, hilaire2024, wein2024}. As such, the present study sets the ground to head toward resource-eﬃcient fault tolerant optical quantum computing. 

\textit{Note} -- During the writing of this manuscript, we became aware of the following works \cite{laneve2024} and \cite{strobel2024} of relevance to the present study. 

\begin{acknowledgments}
 This work was partly supported the European Union’s Horizon 2020 FET OPEN project PHOQUSING (Grant ID 899544), the European Union’s Horizon 2020 Research and Innovation Programme QUDOT-TECH under the Marie Sklodowska-Curie Grant Agreement No. 861097, by the Horizon-CL4 program under the grant agreement 101135288 for the EPIQUE project, by the Plan France 2030 through the projects ANR-22-PETQ-0013, the French RENATECH network, the Paris Ile-de-France Region in the framework of DIM SIRTEQ.

\end{acknowledgments}
\bibliography{biblio}

\end{document}